\documentclass{vgtc}                          %

\ifpdf%
  \pdfoutput=1\relax                   %
  \usepackage{graphicx}                %
  \DeclareGraphicsExtensions{.pdf,.png,.jpg,.jpeg} %
\else%
  \usepackage{graphicx}                %
  \DeclareGraphicsExtensions{.eps}     %
\fi%

\graphicspath{{figures/}{pictures/}{images/}{./}} %

\usepackage{microtype}                 %
\PassOptionsToPackage{warn}{textcomp}  %
\usepackage{textcomp}                  %
\usepackage{mathptmx}                  %
\usepackage{times}                     %
\usepackage{cite}                      %
\usepackage{tabu}                      %
\usepackage{booktabs}                  %
\usepackage{enumitem}
\setlist{nolistsep}

\usepackage[draft]{changes}

\onlineid{3040}

\vgtccategory{Research}

\vgtcinsertpkg

\title{Multi-Focus Querying of the Human Genome Information \\ on Desktop and in Virtual Reality: an Evaluation}

\author{Gunnar Reiske\thanks{e-mail: reiskegw@vt.edu}\\ %
    \parbox{1.4in}{\scriptsize \centering Virginia Tech} %
\and Sungwon In\thanks{e-mail: sungwoni@vt.edu}\\ %
    \parbox{1.4in}{\scriptsize \centering Virginia Tech}
\and Yalong Yang\thanks{e-mail: yalong.yang@gatech.edu (corresponding author)}\\ %
    \parbox{1.4in}{\scriptsize \centering Georgia Tech}} %

\teaser{
  \centering
  \includegraphics[width=\linewidth, height=220pt]{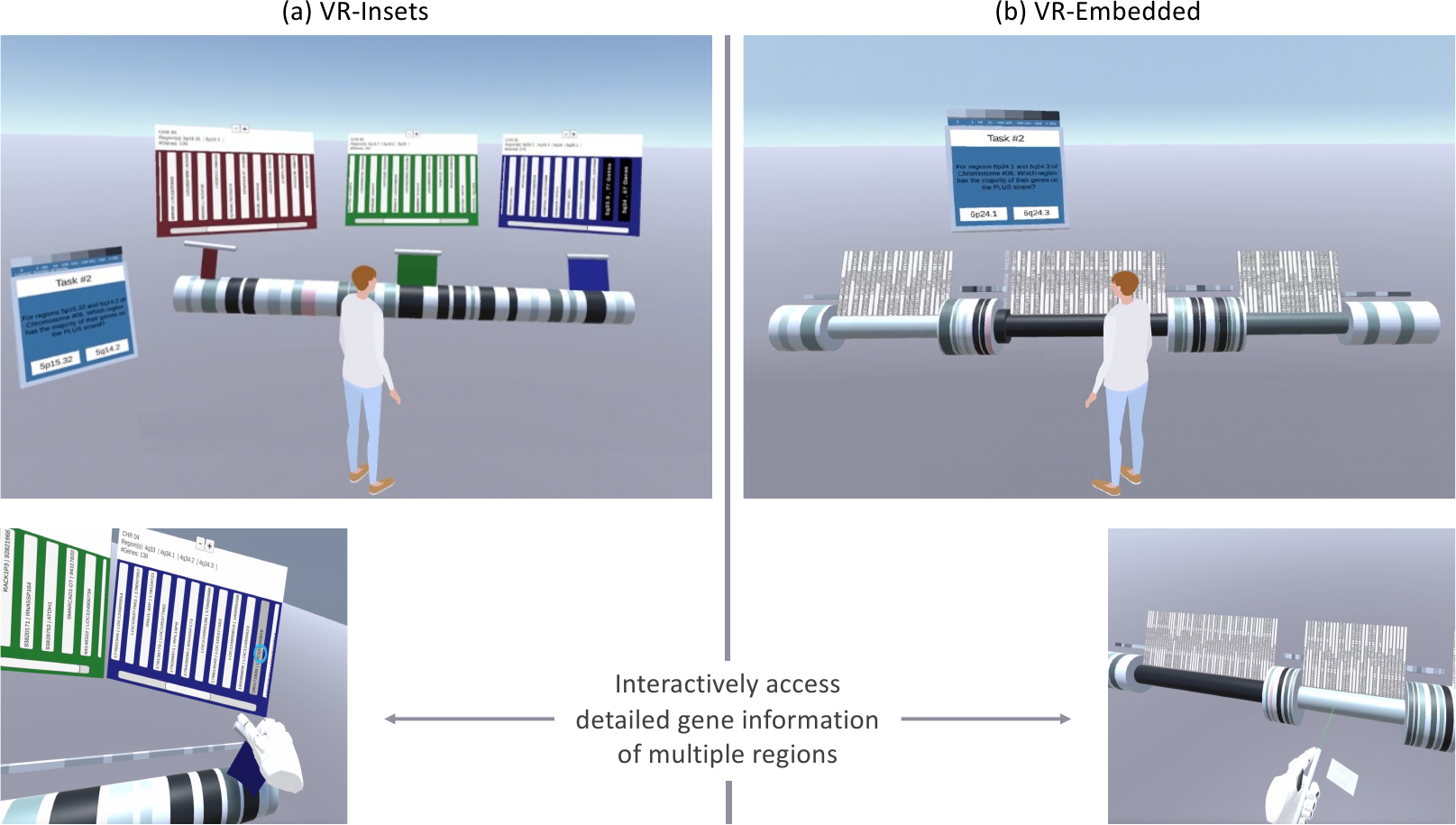}
  \caption{\color{black} Two interaction methods designed for visualizing and interacting with gene information. a) VR-Insets, where the users can see the details of selected genome regions in floating windows; and b) VR-Embedded, where the users can fold and unfold multiple genome regions to switch between different information scales.}
  \label{fig:teaser}
}

\abstract{
The human genome is incredibly information-rich, consisting of approximately 25,000 protein-coding genes spread out over 3.2 billion nucleotide base pairs contained within 24 unique chromosomes. 
The genome is important in maintaining spatial context, which assists in understanding gene interactions and relationships. However, existing methods of genome visualization that utilize spatial awareness are inefficient and prone to limitations in presenting gene information and spatial context. 
This study proposed an innovative approach to genome visualization and exploration utilizing virtual reality. To determine the optimal placement of gene information and evaluate its essentiality in a VR environment, we implemented and conducted a user study with three different interaction methods. Two interaction methods were developed in virtual reality to determine if gene information is better suited to be embedded within the chromosome ideogram or separate from the ideogram. 
The final ideogram interaction method was performed on a desktop and served as a benchmark to evaluate the potential benefits associated with the use of VR. 
Our study findings reveal a preference for VR, despite longer task completion times. 
In addition, the placement of gene information within the visualization had a notable impact on the ability of a user to complete tasks. Specifically, gene information embedded within the chromosome ideogram was better suited for single target identification and summarization tasks, while separating gene information from the ideogram better supported region comparison tasks.%
} %

\CCScatlist{
  \CCScatTwelve{Human-centered computing}{Human computer interaction (HCI)}{Interaction paradigms}{Virtual reality}
}

\usepackage{subfigure}
\usepackage{array}
\usepackage{booktabs, makecell, tabularx}

\begin{document}

\newcommand{\cPC}{\texttt{Desktop}}
\newcommand{\cVRInsets}{\texttt{VR-Insets}}
\newcommand{\cVREmbedded}{\texttt{VR-Embedded}}

\newcommand{\tIdentification}{\texttt{Identification}}
\newcommand{\tComparison}{\texttt{Comparison}}
\newcommand{\tSummarization}{\texttt{Summarization}}

\firstsection{Introduction}

\maketitle

The complete set of genes that make up a human being has around 25,000 protein-coding genes, which are distributed across 3.2 billion nucleotide base pairs stored in the DNA. 
They are unevenly distributed across 22 chromosomes and two sex chromosomes, X and Y, resulting in a total of 24 distinct chromosomes. 
In the field of genetics, there has been a long-standing quest to develop effective methods for visualizing and navigating through the genome to extract gene information and understand their relationships~\cite{nielsen2010visualizing}.

Given the vast amount of information stored in genes, presenting every detail in a comprehensive manner is impractical. Instead, a common approach in genome visualization is to adopt an overview+detail design~\cite{cockburn2009review}. 
An example of this design is the Genome Data Viewer developed by the National Institutes of Health~\cite{rangwala2021accessing}. 
In this viewer, an overview of the genome is presented using ideograms, and users can select specific regions of the ideogram to query and explore detailed information in a separate window. 
While this traditional design is well-suited for focused queries, it poses challenges when users need to compare multiple pieces of information.

To overcome this significant limitation, researchers have proposed multi-focus interactions to enable querying information from multiple sources and visually examining them together~\cite{shoemaker2007supporting,elmqvist2008melange}. 
One commonly employed multi-focus technique is the use of insets. Insets are additional windows that display detailed information from selected portions of the overview~\cite{ghani2011dynamic,lekschas2019pattern}.
However, dealing with multiple insets on a desktop screen can be cumbersome due to the limited physical size of the screen. 
Exploring the potential of leveraging the expansive display space offered by virtual reality (VR), some initial investigations have been conducted, e.g., for geographic maps~\cite{satriadi2020maps} and labels~\cite{lin2021labeling}. 
Nonetheless, there has been a lack of empirical comparisons between insets on desktop screens and those in VR. 
Hence, our first goal is \textit{to provide empirical evidence regarding the effectiveness of insets on both desktop screens and VR platforms}.

However, there is a noticeable drawback when using insets for multi-focus interaction, which is the spatial separation between the overview (ideogram in the case of genome visualization) and the insets. 
As an alternative approach, we propose an embedded design where the insets are spatially integrated into the ideogram itself. 
Notably, not all embedded insets will always be visible, and we should provide users with the ability to interactively show or hide them as needed. 
To achieve this, we have adapted the space-folding technique, allowing for multi-focus querying of genome information~\cite{elmqvist2008melange}: users can unfold a section of the ideogram to reveal its details and fold the detail views back into a compact ideogram form. 
However, when unfolding multiple parts, the linear nature of the genome and ideogram may result in lengthy presentations, requiring more navigation effort from the viewer. 
Embedded insets, compared to standard insets, reduce the cost of context-switching but increase the navigation effort. 
Therefore, our second goal is to \textit{explore this trade-off between embedded and spatially-separated insets in the context of querying genome information.}

The rationale behind the utilization of VR for both embedded insets and standard insets originates from the sheer magnitude and complexity of genome data. 
The large display space~\cite{yang2020embodied} possible in VR allows more genome information to be visualized while providing an innate spatial context. 
The physical selection and navigation in VR~\cite{yang2020embodied,huang2023embodied} provide a more intuitive and effective way to switch between different information levels of the genome~\cite{in2023table,yang2020tilt}. 
Through presenting the genome in VR, users are provided with an innate sense of scale and distance within the genome.

To address the research gaps and obtain empirical evidence for our two research goals, we conducted a user study that compared three different techniques for querying genome information: \cPC{}, \cVRInsets{}, and \cVREmbedded{}.
We excluded the use of embedded insets on the desktop since displaying multiple unfolded parts of the ideogram would exceed the screen size. 
From the genomic~\cite{nusrat2019tasks} and visual information query~\cite[Chap~3]{munzner2014visualization} task taxonomy, we derived three essential gene information query tasks: identifying gene distribution, comparing gene orientation, and summarizing gene-phenotype.
Our findings revealed that participants expressed greater confidence when using VR compared to the desktop environment, although their accuracies were similar across all conditions. \cVREmbedded{} was overwhelmingly preferred over the other two conditions. Furthermore, we observed that embedding gene information within the ideogram was particularly effective for identification and summarization tasks. While having gene information separated from the ideogram better-facilitated comparison tasks.

Our contributions can be summarized in two parts: firstly, we designed and implemented three interactive techniques for multi-focus querying on both desktop and VR platforms, and secondly, we conducted a user study to explore the trade-offs involved in the three developed techniques, namely VR \textit{vs.} desktop and embedded \textit{vs.} spatially separated insets.

\section{Background}

Biological data is often prone to containing large, deeply interconnected sets of data. This is particularly relevant to the field of genetics, where the datasets can be viewed as a series of building blocks that build up into the next one.
In genetics, what defines you as a person, from eye color to food preferences, stems from the patterning of nucleotides that make up your strands of DNA. Every person, even identical twins, has a unique pattern of nucleotide base pairs within their DNA. Specific segments of these patterns, known as genes, are responsible for transcribing proteins that, in turn, express a certain characteristic or trait.

The total sum of genes that defines a human being is estimated at approximately 25,000 spread out over 3.2 billion nucleotide base pairs. Given the considerable size of a person's DNA, nature has devised a way to effectively store the DNA. A single molecule of DNA, which can be hundreds of thousands of nucleotide base pairs long, is tightly wound around special proteins called nucleosomes into long, threadlike structures known as chromosomes. In total, the 3.2 billion base pairs are unevenly divided across 22 chromosomes and the two sex chromosomes X and Y.

These chromosomes are a common state in which genetic data is presented in genome visualizations through what are called ideograms, as each chromosome can easily represent thousands of genes. A long-time pursuit within genetics is the capability of being able to effectively visualize and interact with the genome to parse out gene information and their relationships.

\section{Related Work}

\textbf{Visualizing and interacting with gene information.}
How the genome visualization treats spatial context often determines how much and the degree of detail possible for gene information. 
The primary advantage of having a spatially linked representation is the ability to intuitively determine the location and context of genes within the overall genome. 
Regarding chromosomal representation, they can range from highly realistic 3D representations within VR to a stylized ideogram. 
Realistic depictions have the benefit of providing highly detailed representations of gene regions and features that allow for in-depth analysis~\cite{ABySS, GMOL, arndt2011genome3d}. 
However, given the large size of the human genome, only individual segments can be seen at a time. 
For more stylized depictions, such as ideograms, at the cost of visualizing gene orientation which is the DNA strand that the gene originates from, ideograms have the benefit of showing detailed reports for large segments of the genome at a time. Examples include visually mapping gene locations for a phenotype or disease across the entire genome~\cite{HighRes, Phenogram}. Another benefit, and why ideograms appear to be more popular than realistic representations, is the ability to easily map gene relationships across genomic regions~\cite{Gremlin, Variant}. 
An advantage of utilizing VR for genome visualization is not being constrained by space which can be leveraged to present immersive experiences when navigating through the genome. Existing methods of visualizing and interacting with the genomic dataset in VR typically provide 3D models of genome structures such as chromosome ideograms with which users could interact~\cite{3DGV, 3D, BioVR, 3Dgenomic}.

As mentioned, the human genome has approximately 25,000 protein-coding genes distributed across the genome. Specialized methods of dataset interaction to aid in navigation are imperative to ensure an effective process of analyzing data. 
One popular navigation method for genome analysis can be found in genome browsers. Navigation of the genome dataset in genome browsers works through the application of panning and zooming to chromosome ideograms. 
The ideograms serve a dual purpose. First, they tell the user their exact location within the chromosome. Second, by panning/highlighting regions of the chromosome ideogram, users can zoom into regions of interest~\cite{OrthoZoom}.  Users are able to modify the size of the panning tool to influence to amount of gene information being shown. 
A popular genome browser that utilizes this method of interaction for genome navigation is the UCSC Genome Browser from the University of California, Santa Cruz~\cite{Browser, Browser2}. Another popular genome browser is Genome Data Viewer (GDV) by the National Center for Biotechnology Information~\cite{rangwala2021accessing}. 
A limitation of these genome browsers is that users are limited to a single focus and can only pan over a single region at a time. Another limitation is the repetitive nature of the interaction method requiring multiple pan and zoom actions to move through the dataset.

				\begin{figure*}
						\centering
						\includegraphics[width=\linewidth]{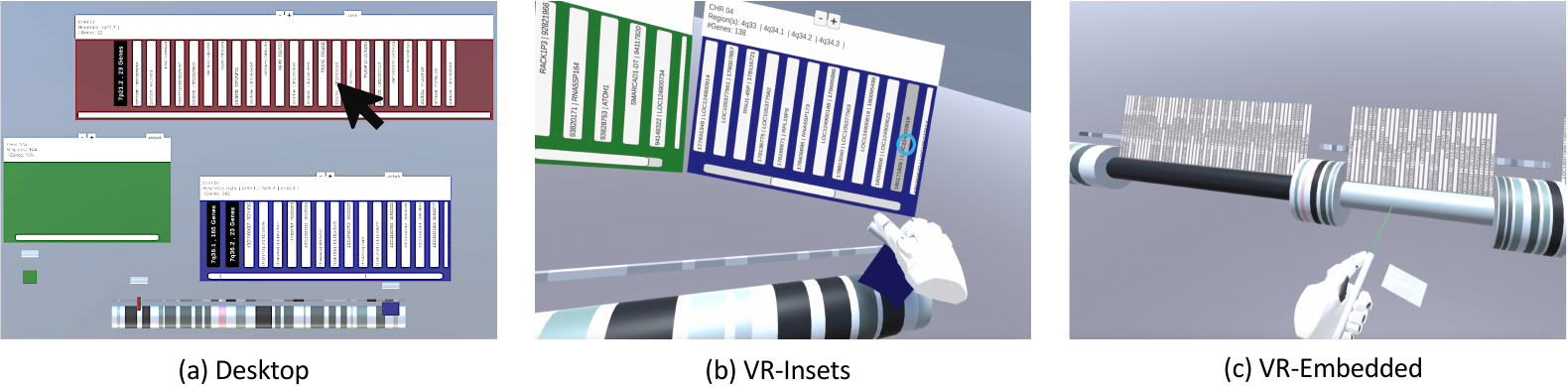}
						\caption{The interaction methods for querying genetic information within the ideogram. (a) \cPC{} uses a mouse pointer for selection, (b) \cVRInsets{} and (c) \cVREmbedded{} use a space-tracked controller to select regions for details in the ideogram. }
                            \vspace{-2mm}
						\label{fig:ideogramInteractions}
					\end{figure*}

\textbf{Multi-focus interaction on desktop and in VR.}
Not all analysis tasks for the human genome are created equal. 
Some tasks deal with viewing a single region while others compare multiple regions~\cite{nusrat2019tasks}. When a task requires regions that are spatially far apart to be compared, a single focus is not efficient. This has led to the development of genome visualizations capable of being multi-focus to maintain context and distance awareness within the genome. 

An interaction method for navigation that supports multi-focus can be found in Melange~\cite{elmqvist2008melange}. The method utilizes space-folding to quickly move across large bodies of data and compare regions that are far apart. 
Space-folding operates on the concept of hiding data between two locations to bring them closer together by ``folding away'' irrelevant data. The degree of space-folding can be modified to manipulate the amount of desired information to be shown and aid in multi-focus approaches. 
A few recent works involve the genome making use of space-folding in some form to modify the amount of genomic information being visualized~\cite{multiscale, HiPiler}.
Variations of this method are quite popular in spatial genome visualizations considering that the genome is linear in its 3.2 billion nucleotide length. 
An example is the use of multi-scale to separate levels of genomic information. 
The genome can be viewed at differing levels of scale through the context of nucleotide patterns that make up genes which make up chromosomes which make up the human genome. 
Through navigating across the levels of scale, genomic detail can be precisely manipulated~\cite{multiscale, scaleTrotter}.

Another method to support multi-focus interaction while still maintaining context and spatial awareness is using multiple windows to view the details. The use of multiple viewing windows was found to aid comparison tasks when complex data was involved~\cite{workingMemory, visualComparisons}. 
Multiple viewing windows were also found to support immersive spaces within VR~\cite{satriadi2020maps}. 
This concept of viewing can be further applied to insets, with previous studies showing that dynamic insets aid in graph navigation while maintaining spatial context~\cite{ghani2011dynamic}. 
Another study found that the use of insets can speed up the visual search while improving comparison accuracy ~\cite{lekschas2019pattern}.

\section{Design and Implementation}

The primary focus of our design strategy was developing and evaluating an innovative method of genome visualization and interaction. Our design prioritized using chromosome ideograms as an overview,
as ideograms can provide an innate spatial context of where the user is within the overall genome. 
A secondary design goal came from evaluating how the placement of genomic information in relation to these ideograms would impact information analysis. The designs were heavily influenced by the biotechnology domain experience and knowledge from the first author.

\textbf{Chromosome Ideogram Design.}
To allow users to effectively explore the human genome without losing spatial context, the genes needed to be accurately mapped to each region of the chromosome ideogram. Before discussing the design process for the chromosome ideograms, some basic background knowledge is required. 
A chromosome is composed of two arms, the \textit{short arm p} and the \textit{long arm q}, which are joined at what is called the centromere to create a complete chromosome.
Chromosome regions are determined through the usage of special stains that create bands of varying darkness and sizes along the arms. With each band representing a region, the regions are named based on the chromosome, arm, and distance from the centromere.

When designing the ideograms, it was a necessity that each chromosome was accurately scaled based on the number of nucleotide base pairs it contained. This also included every region of the chromosome being scaled correctly down to the last nucleotide. Failure to do so would lead to genes appearing at incorrect locations along the chromosome, which is counterintuitive to providing users with accurate spatial context. 
The chromosome ideograms were based on the most recent GRCH38.p14 genome assembly from the National Center for Biotechnology Information. Each chromosome ideogram was carefully mapped to accurately portray the correct region distribution and size. In addition, the coloration for the regions of the ideograms were based on the GRCH38.p14 genome assembly. To aid users in visually identifying regions of high gene activity, a bar that represented gene count could be found above each region. The bar had six possible colorations where the darkness of the bar denoted the approximate gene count for that region, with darker coloration being associated with a higher gene count. 

\textbf{Ideogram Interaction Design.}
The design rationale behind each of the three ideogram interaction methods can be summarized as:
\begin{itemize}[leftmargin=*]
    \item \textit{Precise position indication.} The method should consistently indicate the user's exact position within the genome, ensuring they are aware of their location at all times.
    \item \textit{Controllable gene visualization.} Users should have the ability to fine-tune the number of genes they wish to visualize, enabling them to focus on specific gene sets according to their needs.
    \item \textit{Multi-focus support.} The method should facilitate multi-focus functionality, allowing users to access and simultaneously view spatially distant chromosome regions. 
    \item \textit{Multivariate gene visualization.} In addition to presenting the exact positions of genes within the chromosome, the method should be capable of visualizing other relevant gene features, such as gene orientation and gene phenotypes.
\end{itemize}

In summary, we include two interactions methods in our study.
The first interaction method, \cVREmbedded{}, had all gene information embedded within the ideogram. 
Conversely, the \cVRInsets{} interaction method displayed the gene information at a certain distance from the ideogram. 
Due to VR being capable of providing excellent spatial context, significant emphasis was placed on developing interaction methods within VR. 
To evaluate the impact of VR, we also included a \cPC{} interaction, which was an interpretation of the \cVRInsets{} interaction method. 
     
\begin{figure}
						\centering
						\includegraphics[width=\columnwidth, height=0.5\linewidth]{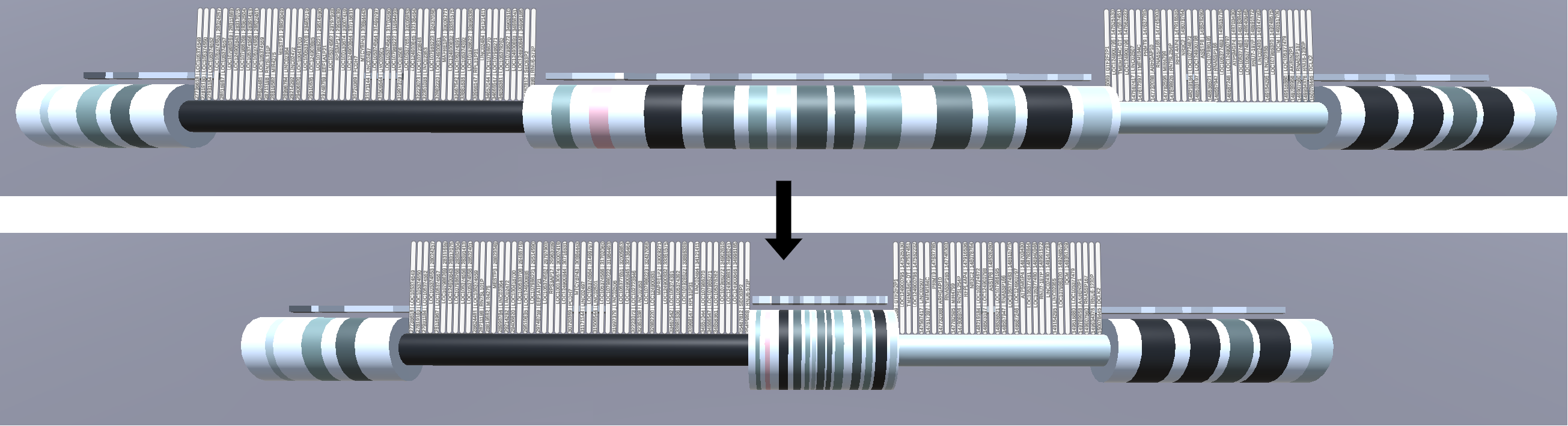}
						\caption{Interactively compressing a region in the chromosome.} 
						\label{fig:wandCom}
                            \vspace{-2mm}
					\end{figure}

\textit{\cVREmbedded{}.} 
The gene information was directly embedded within the chromosome, as shown in \autoref{fig:ideogramInteractions} (c).
To access the gene information, a specialized tool referred to as the wand was used. 
Insertion of the wand into the chromosome allowed users to select specific regions of the ideogram to \textit{open}, \textit{close}, or \textit{compress}. 
Working directly on the chromosome ideogram allowed users to intuitively know their exact position within the genome at all times. 
Users were not limited to a single region of the chromosome and could freely manipulate the ideogram as they saw fit. This meant that they could fine-tune the number of genes being visualized at any point in time and utilize a multi-focus approach to view regions that are spatially far apart by compressing regions as shown in \autoref{fig:wandCom}.

When the chromosome was opened to investigate gene information, the user was provided with two points of data for the genes contained within. The genes were ordered based on the starting position of the first nucleotide base pair that coded out the gene. The genes were also oriented based on the DNA strand that the gene originates from. If the gene was located on the plus strand, the gene information was listed as starting base pair followed by the gene symbol. The information was displayed on the ideogram from the bottom up. If the gene was located on the minus strand, the gene information was listed as a gene symbol followed by the starting base pair. The information was shown on the ideogram from the top down. When visualizing phenotypes along the chromosome, the phenotype would be represented by a colored cube. If a region contained a gene with the phenotype, the associated color cube would be located above the region. Upon opening the region, any subsections that had genes with the phenotype would then have the colored cube above. Finally, opening the subsection would reveal the genes of the phenotype, denoted by the colored cube on the gene itself.

\textit{\cVRInsets{}.} 
It is a method of chromosome interactively revealing gene information separate from the ideogram.
\cVRInsets{} was composed of two parts: the viewing windows and the scope. 
The scope and ideogram interaction worked on the same principle as the wand tool in \cVREmbedded{}. 
Insertion of the scope into the ideogram allowed users access to the regions where the scope intersected. 
However, instead of the gene information appearing directly on the chromosome, it was populated on a moveable window separate from the ideogram as shown in \autoref{fig:ideogramInteractions} (b). 
The user could freely resize the scope and viewing windows, which moderated the quantity of gene information displayed in the viewing windows. 

A multi-focus approach was achieved by allowing users to have multiple viewing windows at a time to view different regions of the ideogram concurrently. When gene information was presented on the viewing window, the gene ordering and orientation followed the same format as \cVREmbedded{} interaction. When visualizing phenotypes of the chromosome ideogram, the phenotype would be represented by a colored cube on the ideogram similar to \cVREmbedded{} interaction. When the region was intersected by the scope and loaded to the window, the window populated with the region name, number of genes, and phenotypes found inside the region. Upon opening a region containing phenotype-mapped genes, the phenotype genes were denoted by the color associated with them.

\textit{\cPC{}.} This condition is the implementation of \cVRInsets{} as a desktop application. 
\cPC{} interaction operated the same as \cVRInsets{} with only two exceptions. The first was that users operated the method via mouse input instead of through controllers in VR. The other exception was that the windows had a toggling lock option that could lock windows in place. This was done to prevent misclicks within the limited workspace. The layout of the \cPC{} method can be seen in \autoref{fig:ideogramInteractions} (a).
\section{User Study}
The main aim of this work is to understand the trade-offs in using different interaction techniques for querying human genome information, particularly between different computing environments (desktop and VR) and different insets placement strategies (spatially separated and embedded).
To this end, we designed and conducted a user study to compare three interaction techniques.

\textbf{Experiment Setup.}
All interaction methods were implemented using Unity 3D. 
A Meta Quest 2 headset with controllers for interaction was used for the two VR conditions. The VR headset was connected to a PC with a RTX 3070 graphics card for optimal rendering performance.
The desktop condition was tested on a 15.6-inch screen with 1920$\times$1080 resolution with a mouse for interaction.

\textbf{Participants.} 
15 participants (ten male and five female) were recruited from students at a university: 11 were graduate students, and 4 were undergraduate students. 
The dominant major was CS at 11, with the remaining 4 being a mix of biology-related majors and majors adjacent to CS. The age of the participants ranged from 20 to 30 years old. To be eligible for the study, participants were required to be 18 years or older and have basic knowledge of biology.
At the completion of the study, each participant was compensated for their time and effort with a \$20 Amazon gift card.

\textbf{Dataset and Tasks.}
The study data were from the National Center for Biotechnology Information Homo Sapiens Annotation~\cite{NCBIdata}.

In order to evaluate interactive querying methods, we carefully selected tasks that required users to navigate and visually inspect genome information in a significant manner. 
Additionally, these tasks were designed by the first author, a domain expert in biotechnology, to encompass three essential features for understanding and analyzing genomes~\cite{nusrat2019tasks}. 
These features consist of the distribution of genes across the chromosome, the orientation of genes within chromosome regions, and the phenotypes associated with the genes on the chromosome.
Meanwhile, to effectively capture the visual query perspective in relation to genome information, we adopted three visual query actions based on Munzner's taxonomy~\cite[Chap~3]{munzner2014visualization}.
We aligned the three key genome features with the corresponding visual query actions, resulting in the formulation of the following study tasks:

				\begin{figure*}
						\centering
						\includegraphics[width=\linewidth]{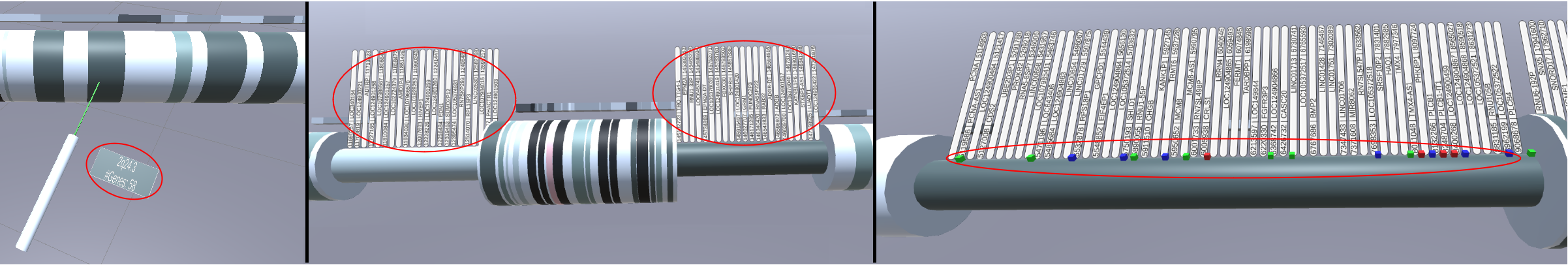}
						\caption{Example of each task for \cVREmbedded{}. In the identify gene distribution task, the wand tool was used to skim the ideogram to locate the region with a specific gene count (left). In the compare gene orientation task, participants used the wand tool to open regions for comparison (center). In the summarize gene phenotype task, participants utilized the cubed markers representing phenotypes mapped along the ideogram to determine the dominant phenotype (right). } 
                            \vspace{-2.5mm}
						\label{fig:embeddedTasks}
					\end{figure*}

\textit{Identify gene distribution.} 
Genes are not evenly distributed across chromosomes. Instead, they are found in clusters with varying sizes. This results in overall chromosome size being deceiving as large regions of a chromosome can be gene sparse and vice versa. For example, chromosome 11 is approximately 70\% the size of chromosome 04. However, it has over 400 more genes. 
When tasks were given, participants were presented with a series of ideograms and asked to identify the region with a specific gene number for each chromosome. The gene count bar associated with the ideogram provided additional context for gene distribution. A demonstration of the task using \cVREmbedded{} is provided in \autoref{fig:embeddedTasks}.

\textit{Compare gene orientation.}
DNA is double-stranded, and the strand denotes the orientation of a gene. For example, genes found on the 5' to 3' strand are given a plus orientation, while genes on the complementary 3' to 5' strand are denoted with a minus orientation. Understanding genes' spatial distribution and orientation is of critical interest in understanding gene interactions, as genes on the same orientation have a higher likelihood of interaction. 
When tasks were given, participants were again presented with a series of ideograms and asked to compare the gene orientation of two given regions. The regions had to be located, and their gene orientation visually inspected for comparison. From the regions, they were asked to specify whether the gene is dominated by a plus or minus gene orientation. An example using \cVREmbedded{} is provided in \autoref{fig:embeddedTasks}.

\textit{Summarize gene phenotype.} 
The vast majority of human phenotypes are classified as polygenic traits. This means that multiple genes are involved in producing the phenotype. For example, polygenic traits include height, eye color, and hair color. 
In addition, complex diseases such as Alzheimer's and Schizophrenia are also the result of multiple genes interacting. 
Genes associated with these polygenic traits and complex diseases can be found scattered along a single chromosome or throughout the genome. 
The final task was to highlight the capability of the interaction methods to aid in exploring the phenotype-mapped chromosomes. A mapped chromosome with the three imaginary phenotypes A, B, and C was given to participants. Using the provided interaction method, participants were asked to summarize the dominant phenotype of the chromosome. A demonstration of the task using \cVREmbedded{} is provided in \autoref{fig:embeddedTasks}.

Furthermore, to ensure fairness in all task trials, we enforced a restriction that each  ideogram could only be used once throughout the entire study. This measure was implemented to minimize the learning effect. Moreover, we classified the chromosomes into three groups based on their similarity to eliminate any potential influence of chromosome differences on the collected measures.

To evaluate the feasibility of the procedure and tasks, particularly to ensure people with basic biology background can understand and complete the tasks, we piloted our study tasks with three computer science graduate students. 
All three pilot participants could finish the study without struggle.
Based on the pilot, we anticipated that our study tasks are generic information query tasks and a significant amount of biology knowledge was not required from the participants.

\textbf{Study Procedure.}
The study had three main components, with an average completion time of approximately 90 minutes. 
Each participant completed 18 study trials:
3 interaction conditions $\times$ 3 study tasks $\times$ 2 repetitions. 
To reduce potential learning bias, the order of the interaction methods followed a Latin Square design.

\textit{Introduction.} At the beginning of the study, each participant received a brief overview of the procedure of the study and signed the consent form. They were then presented with a background questionnaire to collect their demographic information. To ensure all participants understood the context of why chromosome ideograms were visualizing the human genome, a brief lesson was provided. The lesson covered genomics basics, including DNA, genes, and how they are stored within chromosomes. 

\textit{Training and study task.} The first time a participant was to use an interaction method, they were required to view a training video that covered its features. 
Afterward, they were provided with a training environment to familiarize themselves with the interaction method. This was repeated for each interaction method. Once the conductor confirmed familiarity with the interaction method, participants were required to fully complete a task. Participants were asked to complete two repetitions for each task. 
At the conclusion of each task across all conditions, a mandatory 5-minute break was given to prevent participant fatigue.

\textit{Questionnaires.} Following the completion of a task for each interaction method, the participant was asked to score their confidence in using the interaction method to complete the task. The scoring was from one to five, with five denoting the highest level of confidence. Participants were given an exit survey after completing all tasks and questionnaires. The survey covered their strategy behind using the interaction methods, any complications they experienced using the interaction methods, and ranking the methods based on preference in completing the tasks. 
After the exit survey, participants were given a modified version of the NASA-TLX based on a seven-point scale to assess the perceived workload of the three interaction methods.

\textbf{Measures.}
We collected the performance and interaction data to obtain a nuanced understanding of  user experience when possible.

\textit{Performance Measures.} For all three tasks the \textit{completion time}, \textit{accuracy}, and \textit{perceived confidence} were collected. 
Completion time was measured from when the participant started the task to when they submitted their answer. 
Accuracy was collected based on the correctness of the participant and marked as either correct or incorrect. 
Perceived confidence was collected at each task's end and presented on a five-point Likert scale. 

\textit{Visual Querying Breakdown Measures.} 
For the \textit{identify gene distribution} task, participants were instructed to query the chromosome ideogram until the region with a specific gene count was identified. From this, measures that collected the percentage of the chromosome that had to be queried and how long it took participants to first identify the region were collected. 
In the \textit{compare gene orientation} task, participants were given two regions to locate on the ideogram and compare the orientation of their genes. 
From this, measures that collected the percentage of the chromosome that had to be queried and how long it took participants to compare the two given regions and come to an answer after locating both regions were collected. 
For the \textit{summarizing gene phenotype} task, participants did not need to query the chromosome to locate the regions as they were already marked by the presence of the phenotype-mapped genes. Thus, we did not collect its breakdown data.

\section{Results}
For the measures that meet the normality assumption, repeated measures ANOVA was used. Measures that fell under this category were completion time, locating region time, analysis time, and chromosome exploration percentage. 
Tukey’s HSD post-hoc tests were performed to allow for pairwise comparisons of the measures returned by the three interaction methods. 
For the measures that were unable to meet the normality assumption, the Friedman test was used for analysis. These measures were task accuracy, interaction method confidence, and NASA-TLX scores. The Nemenyi post-hoc test was used for the pairwise comparisons. 
Significance values of $p < 0.05(*)$, $p < 0.01(**)$, and $p < 0.001(***)$ are reported using the number of stars in the parenthesis. If the significant value fell between $0.05 \leq p \leq 0.1$ it was reported as marginally significant. 

\subsection{Performance}
    
\begin{figure}
						\centering
						\includegraphics[width=\columnwidth]{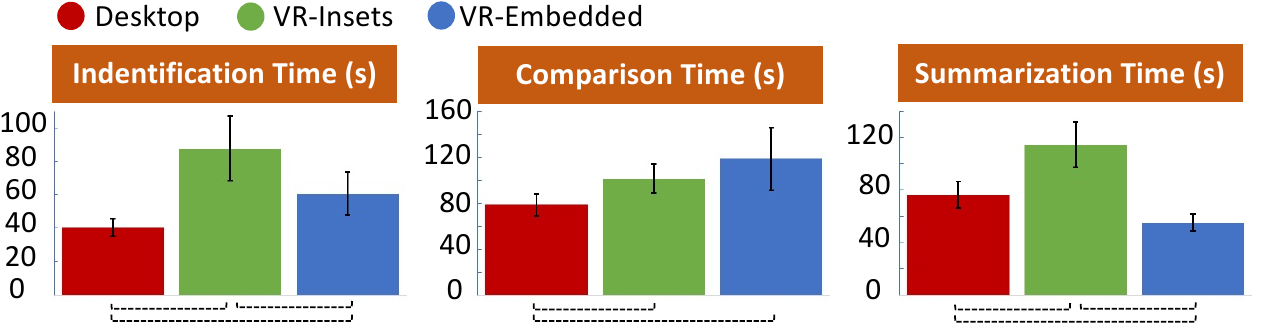}
						\caption{Average completion time of each task for each interaction method. The confidence intervals indicate 95\% confidence. A dashed line indicates p \textless 0.05.} 
						\label{fig:compTime}
					\end{figure}

\textbf{Completion Time.}
The average completion time for all tasks can be found in \autoref{fig:compTime}. For the identify gene distribution task, \cPC{} (avg 40.3s, CI=5.12s) was significantly faster than \cVREmbedded (avg 60.68s, CI=13.18s, $*$) and \cVRInsets{} (avg 87.62s, CI=19.35s, $***$). \cVRInsets{} was also significantly slower than \cVREmbedded{} ($*$). 
In the compare gene orientation task, \cPC{} (avg 78.39s, CI=9.65s) was significantly faster than \cVRInsets{} (avg 101.36s, CI=12.49s, $*$) and \cVREmbedded (avg 118.5s, CI=27.24s, $**$). 
For the summarize gene phenotype task, \cVREmbedded (avg 54.88, CI=6.56s) was significantly faster than \cPC{} (avg 76.47s, CI=10.16s, $*$) and \cVRInsets{} (avg 114.09s, CI=17.13s, $***$). \cVRInsets{} was also slower than \cPC{} ($***$).

\textbf{Accuracy.}
Task accuracy across all three tasks for the interaction methods was either close to 100\% or 100\% accuracy, with variations between interaction methods being one to two incorrect answers. 
For the identification task, \cVRInsets{} (avg 97\%, CI=3\%) had the lowest accuracy while \cPC{} (avg 100\%, CI=0\%) and \cVREmbedded{} (avg 100\%, CI=0\%) had perfect participant accuracy. 
For the comparison task, \cPC{} (avg 90\%, CI=10\%) has the lowest accuracy while both \cVREmbedded{}(avg 93\%, CI=7\%) and \cVRInsets{} (avg 93\%, CI=7\%) were tied at the same accuracy. \cPC{} (avg 93\%, CI=7\%) also had the lowest accuracy. 
For the summarization task, \cVREmbedded{} (avg 100\%, CI=0\%) and \cVRInsets{} (avg 100\%, CI=0\%) had perfect participant accuracy.

\textbf{Perceived Confidence.}     
The majority across all three tasks was a 4 or 5 rating. This led to only slight confidence variations between the three interaction methods across all three tasks. For the identification task, participants had the highest confidence using either \cPC{} (avg 4.73, CI=0.23) or \cVRInsets{} (avg 4.73, CI=0.21) while \cVREmbedded{} (avg 4.67, CI=0.25) had the lowest confidence. 
In the comparison task, participants had the highest confidence using \cVRInsets{} (avg 4.8, CI= 0.17) to complete the task and the least confidence with \cVREmbedded{} (avg 4.6, CI=0.18). The task confidence for \cPC{} (avg 4.63, CI=0.18) was in between. 
Lastly, for the summarization task, participants had the highest confidence using \cVREmbedded{} (avg 4.37, CI=0.26) followed by \cVRInsets{} (avg 4.23, CI=0.29) and \cPC{} (avg 4.13, CI=0.37).

\subsection{Visual Querying Breakdown Analysis}

				    \begin{figure}
						\centering
						\includegraphics[width=\columnwidth]{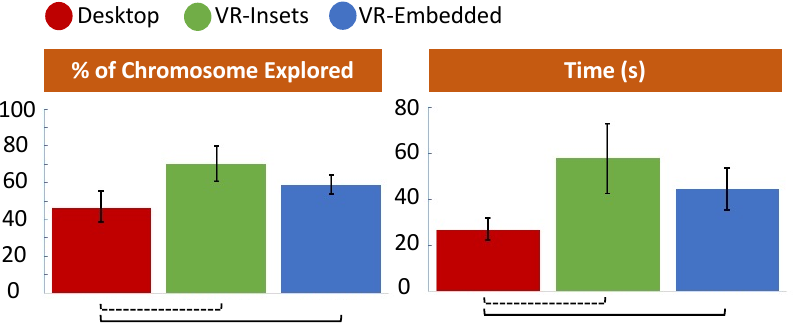}
						\caption{The percentage of chromosomes explored and time taken to identify the specific region in the identification task. The confidence intervals indicate 95\% confidence. A dashed line indicates p \textless 0.05. A solid line indicates marginal significance for  0.05 \textless p \textless 0.1.} 
                            \vspace{-3mm}
						\label{fig:chrIdent}
					\end{figure}

     			\begin{figure}
						\centering
						\includegraphics[width=\columnwidth]{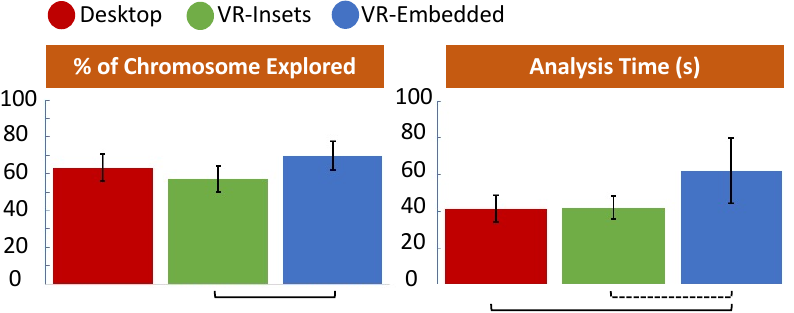}
						\caption{The percentage of chromosomes explored and time taken for analysis in the comparison task. The confidence intervals indicate 95\% confidence. A dashed line indicates p \textless 0.05. A solid line indicates marginal significance for  0.05 \textless p \textless 0.1.} 
						\label{fig:chrCompare}
					\end{figure}

\textbf{Total chromosome explored.} For the identification task, \cPC{} (avg 46\%, CI=8\%) required a marginally significant lesser amount of the chromosome to be queried compared to \cVREmbedded{} (avg 59\%, CI=5\%) and a significantly lesser amount than \cVRInsets{} (avg 70\%, CI=10\%, $***$), \autoref{fig:chrIdent}. In the comparison task, \cVRInsets{}(avg 57\%, CI=7\%) required the least amount of the chromosome to be queried while \cVREmbedded{} (avg 70\%, CI=8\%) required the most. The difference between the two was found to be marginally significant. \cPC{} (avg 63\%, CI=8\%) was in between \cVRInsets{} and \cVREmbedded{}, \autoref{fig:chrCompare}.

 \textbf{Time taken for identifying gene count in a single region.} During the identify gene distribution task, the average time it took a participant to first reach the region with the correct gene count was measured. \cPC{} (avg 26.76s, CI=4.93s) was significantly faster than \cVRInsets{} (avg 57.73s, CI=15.17s, $**$) and \cVREmbedded (avg 44.32s, CI=9.16s, $*$), \autoref{fig:chrIdent}.
  
\textbf{Time taken for gene region comparison.} 
In the compare gene orientation task, for the time it took each participant to compare the genes' orientation, \cPC{} (avg 41.52s, CI=7.10s) was significantly faster than \cVREmbedded (avg 62.10s, CI=17.84s, $*$). \cVRInsets{} (avg 42.04s, CI=6.36s) was faster than \cVREmbedded by a marginal significance, \autoref{fig:chrCompare}.

				    \begin{figure}
						\centering
						\includegraphics[width=0.6\columnwidth]{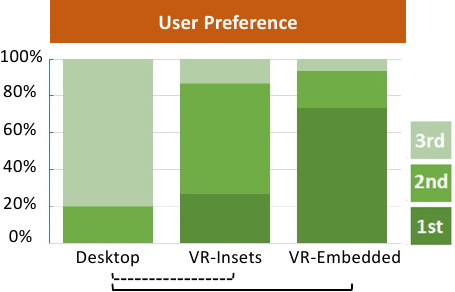}
						\caption{Participant preference and confidence ranking of each interaction method. A dashed line indicates p \textless 0.05.} 
						\label{fig:userConPref}
					\end{figure}

				    \begin{figure*}
						\centering
						\includegraphics[width=\linewidth]{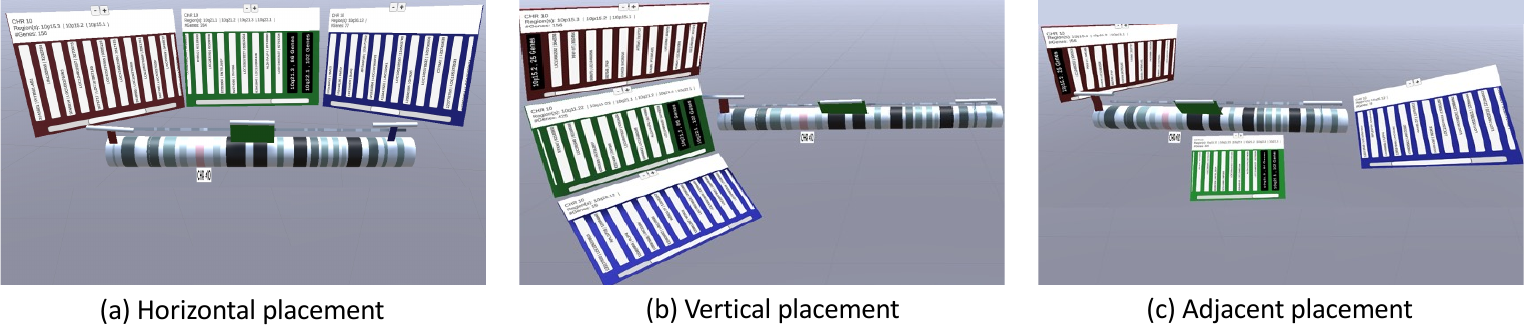}
						\caption{Three layout strategies used by our participants in \cVRInsets{}. (a) participants horizontally placed the insets. (b) participants vertically placed the insets. (c) participants placed insets close to the identified region.} 
                            \vspace{-1.5mm}
						\label{fig:windowConfig}
					\end{figure*}

\subsection{Preference and Perceived Task Load} 
\textbf{Preference.} 
\cVREmbedded{} was the most preferred interaction method in the study with 80\% of the participants. The \cPC{} was selected as the least preferred by 80\% of the participants against \cVRInsets{} ($*$) and \cVREmbedded{} ($**$). The participant preference rankings can also be found in \autoref{fig:userConPref}. 

\begin{figure}
						\centering
						\includegraphics[width=0.85\columnwidth]{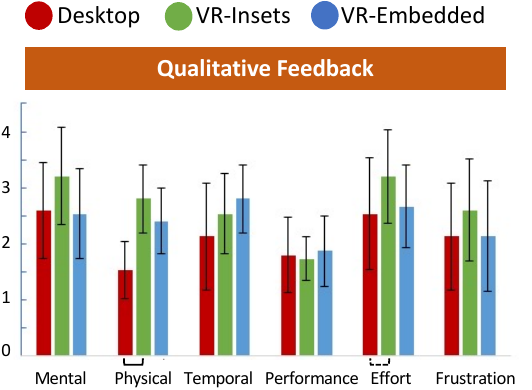}
						\caption{NASA-TLX score for each interaction method. NASA-TLX modified to a seven-point scale. Lower values denote a better experience. The confidence intervals indicate 95\% confidence. A dashed line indicates p \textless 0.05. A solid line indicates marginal significance for  0.05 \textless p \textless 0.1.} 
                            \vspace{-3mm}
						\label{fig:nasa}
					\end{figure}

\textbf{Perceived Task Load.} 
For the NASA-TLX, a lower score denoted a better experience. \cPC{} (avg 1.53, CI=0.51) was considered the least physically demanding. \cVREmbedded{} (avg 2.4, CI= 0.58) was perceived to be significantly less physically demanding than \cVRInsets{}(avg 2.8, CI=0.6, $*$). Self-perceived user performance was relatively equal across the three interaction methods. \cVRInsets{} (avg 3.2, CI=0.84) was identified as requiring the most effort with marginal significance over \cPC{} (avg 2.53, CI=1.00), which required the least. \cVRInsets{} was also considered to be the most frustrating interaction method. Mental demand was relatively close between \cPC{} (avg 2.6, CI= 0.86) and \cVREmbedded{} (avg 2.53, CI=0.81), while \cVRInsets{} (avg 3.2, CI=0.87) polled as having highest mental demand. For temporal demand, \cVREmbedded{} (avg 2.8, CI=0.60) returned as the highest and \cPC{} (avg 2.13, CI=0.96) as the lowest.
The ratings are demonstrated in \autoref{fig:nasa}.

\subsection{Participant Interaction Strategies}
At the end of each task in the study, the final workspace setup was recorded. Participant was also asked to elaborate on their strategies.

In general, the process of setting up their workspace was relatively consistent. When setting up the chromosome ideogram for interaction in VR, participants preferred the chromosome in a horizontal position. Some participants attempted to position the ideogram vertically but reverted back to a horizontal orientation. When positioning the viewing windows in \cVRInsets{}, we observed that the viewing windows had three organization strategies. The windows were placed side-by-side, stacked vertically, or scattered within the workspace, as shown in \autoref{fig:windowConfig}. When positioning the viewing windows on the \cPC{}, they were placed side-by-side or stacked vertically. 
Our observations also indicate that participants interacted more in \cVRInsets{} than \cPC{}, particularly in terms of resizing the scope, resizing, and moving the viewing windows.

\textbf{Identify gene distribution task.} 
For \cPC{}, nine participants resized the scope when searching for a specific gene count. We observed two strategies for resizing the scope: minimizing to view one region at a time, and maximizing to view all regions at the same time. In \cVRInsets{}, ten participants chose to resize the scope when navigating through the ideogram. Similar to \cPC{} interaction, they resized the scope for the same reason. In \cVREmbedded{}, participants opted for two strategies, they either started on either end of the chromosome until they found the region or used the gene distribution bar to aid them in identifying the region.

\textbf{Compare gene orientation task.} 
For \cPC{}, six participants resized the scope, while seven resized the viewing windows. Furthermore, eight participants re-positioned the windows from their default locations. In \cVRInsets{}, eight participants resized the scope, while ten participants resized the viewing windows. In addition, twelve participants re-positioned one or more of the viewing windows when comparing gene orientation across regions. A common strategy for both \cPC{} and \cVRInsets{} involved resizing the viewing window size to show all the regions in the gene while placing one window above the other to compare gene orientation. In \cVREmbedded{}, six participants used the chromosome compression feature while the rest opened both regions and physically moved between them. 

\textbf{Summarize gene phenotype task.} 
For \cPC{}, six of the participants resized the scope, while eight resized the viewing windows. In addition, nine participants re-positioned the windows from their default locations. In \cVRInsets{}, nine participants resized the scope, while eleven participants resized the windows. Moreover, fourteen participants re-positioned one or more of the viewing windows when comparing gene phenotypes across regions. Similar to the gene orientation task for  and window, a common strategy was maximizing the size of viewing windows and stacking the windows vertically. For \cVREmbedded{}, participants expanded the regions containing the phenotypes and stepped back to take in the entire ideogram.

\subsection{Qualitative Feedback}

\textbf{\cVREmbedded{}.} The most prevalent feedback given pertained to object-grabbing. Multiple participants mentioned minor inconveniences when grabbing objects clustered closely together. One participant stated how \textit{``initially working with VR was a little difficult''} as they would grab the wrong objects. Another commented how it caused interruptions to their workflow as they \textit{``would sometimes grab the wrong object and would have to restart''}.

\textbf{\cVRInsets{}.} Similar to \cVREmbedded{}, multiple participants expressed dissatisfaction with the sensitivity of object-grabbing. One participant noted that when multiple insets were used closely together that it became \textit{``difficult to try to control which one I grabbed''}.

\textbf{\cPC{}.}  The most common response concerned the available working space for \cPC{}. Multiple participants commented on limited working space and wished for a larger space to resize the viewing windows. One participant commented how the \textit{``area wasn't large enough''}, which impacted their strategy. Another stated that \cPC{} was \textit{``hard to use''} for the gene phenotype summarization task due to the limited workspace.

\section{Discussion}

\textbf{Tradeoff for the use of insets in desktop and VR (\cPC{} vs. \cVRInsets{}).} 
\cPC{} interaction consistently outperformed \cVRInsets{} in task completion time. 
Even more, in the gene distribution task, \cPC{} performed better when participants needed to identify the region of a specific gene count and required a smaller portion of the chromosome to be queried (see \autoref{fig:chrIdent}). 
On average, participants only took 26.76 seconds to locate the correct region for \cPC{}, while 57.73 seconds for \cVRInsets{}. In addition, the \cPC{} slightly outperformed \cVRInsets{} in task accuracy and had higher confidence in completing the task. It can be noted, however, tasks requiring a larger portion of the chromosome to be visualized concurrently (multi-focus) \cVRInsets{} outperformed \cPC{}. In the gene orientation and phenotype tasks, \cVRInsets{} experienced a higher task accuracy and confidence.

We observed that the physical movement in VR during the process of setting up the workspace took longer than clicking and dragging on \cPC{}. The qualitative feedback further supports this as participants responded with \cVRInsets{} being the most physical, as shown in \autoref{fig:nasa}. Once the participants finished the setup, the time for interacting with chromosome ideograms between the two interaction methods was close to equal. This aligns well with the comparison time for gene orientation (see \autoref{fig:chrCompare}), with a time difference of less than a second in favor of the \cPC{}.

In summary, \cPC{} had a better completion time due to the lower physical demands and performed better in the task that did not require multi-focus. 
However, participants tended to be slightly more accurate and confident with \cVRInsets{} for the tasks that leaned more on using multi-focus viewing windows.

\textbf{Trade-off between embedded and spatially-separated
insets in VR (\cVREmbedded{} vs. \cVRInsets{}). } 
For the gene distribution task, the \cVREmbedded{} method performed better and had a faster completion time than \cVRInsets{}. In addition, \cVREmbedded{} required less of the  chromosome to be queried, which leads to a faster time in identifying the required gene region  (see \autoref{fig:chrIdent}). This could be potentially due to the gene information being directly embedded into the ideogram. In single target tasks, users could focus on their ideogram in front of them, instead of having to repeatedly shift their gaze to a separate viewing window in \cVRInsets{}.

For the gene-phenotype task, \cVREmbedded{} again outperformed \cVRInsets{}. We observed that users could quickly get an overview of phenotype distribution with embedded and visible phenotype markers along the ideogram while \cVRInsets{} needed to scroll through the viewing window. This was reflected in the \cVREmbedded{} being about a minute faster and giving participants a higher degree of confidence.

For the gene orientation task which required participants to compare two regions far apart, embedded gene information worked against \cVREmbedded{}. In \cVRInsets{} participants could position the viewing windows for the purpose of side-by-side quick comparisons. This was not possible for \cVREmbedded{}, which led to participants having to alternate between the regions during the comparison process and having a longer completion time, analysis time, and required a larger percentage of the chromosome to be queried (see \autoref{fig:compTime} and \autoref{fig:chrCompare}). 

In summary, we found that \cVREmbedded{} performed the best for single target identification and summarization while \cVRInsets{} was the optimal choice for comparison-based tasks.

\textbf{General takeaways.} 
Overall, no single ideogram interaction method performed the best across all three tasks. There was, however, participants' preference towards the use of VR over the \cPC{} to complete the tasks as shown in \autoref{fig:userConPref}. In addition, VR gave participants more confidence in completing tasks over the \cPC{}. 
In the gene distribution task, \cPC{} outperformed the VR interaction methods (see \autoref{fig:chrIdent}). 
This partly carried over to the gene orientation comparison tasks as \cPC{} was the fastest interaction method. This outcome may be due to the \cPC{} requiring less physical movement compared to any of the interaction methods in VR. For tasks where participants needed to take in everything at once, such as summarizing traits in the gene-phenotype task, the VR interaction methods came out on top. This could be due to the additional working space in VR, which allowed participants to look at the entire workspace. This was also reflected in the participants having higher task accuracy and confidence for \cVREmbedded{} and \cVRInsets{} over \cPC{}.

\textbf{Generalizability, limitations, and future work.}
We conducted an investigation into the effectiveness of three multi-focus interaction methods specifically designed for exploring ideograms, which can be regarded as a one-dimensional data structure.
Although our focus was on addressing a specific problem within a particular domain, our findings and implementations offer high-level interaction techniques that can be readily adapted to other applications, particularly those involving data with inherent one-dimensional structures such as timelines or certain nanomaterials. 
Moreover, there is potential to extend the use of our techniques to data with additional dimensions, such as two-dimensional geographic maps~\cite{satriadi2020maps}, matrices~\cite{lekschas2019pattern}, or even three-dimensional scientific data. 
However, the process of adapting these techniques may not be straightforward, as the dimensionality can significantly impact the visual representation and interaction dynamics. Fortunately, previous research, such as the work on Melange~\cite{elmqvist2008melange}, has provided valuable insights and experience in developing two-dimensional multi-focus interactions.

During the study, several participants expressed frustration when attempting to perform precise movements in VR. 
This challenge has long been recognized as a drawback of mid-air interaction, prompting the research community to invest significant efforts in addressing this issue. Various approaches have been explored to improve precision in VR interactions. 
These include the establishment of clear interaction boundaries to guide users' movements, the inclusion of visual indicators to highlight focused objects and aid in targeting, and even the incorporation of physical proxies that leverage tangibility to enhance precise input and control~\cite{satriadi2022tangible,smiley2021made,cordeil2020embodied}.
Exploring the integration of these aforementioned techniques into our specific problem represents a promising avenue for further research.

An additional limitation brought forward by the participants was the limited working space on \cPC{}. To address this issue, one potential solution would be to provide an infinite canvas as the working space integrated with a pan and zoom interface to allow users to pan in any direction and zoom in and out within the desktop environment. This approach is similar to platforms like Miro and Google Jamboard, which offer expansive virtual workspaces.

In regards to potential future work, modifications to the participant pool and means of interaction with the system would be considered. In the user study, the participant age range was 20 to 30 years old. To better represent the potential user base for the application, a more diverse age range of participants would be necessary for any future work. Other methods of system feedback will also be considered. This includes the use of vibrations when interacting with the system to increase precision and the inclusion of leader lines for \cVRInsets{} to better visualize linking.

Lastly, the \cVREmbedded{} and \cVRInsets{} methods were evaluated separately in our work.
However, both methods have the potential to operate concurrently on the same chromosome ideogram. 
A viable future work would be combing \cVREmbedded{} and \cVRInsets{} together as a hybrid multi-focus interaction technique.

\section{Conclusion}
In this paper, we have developed and evaluated three different interaction methods for visually querying human genome information. These methods were specifically designed to enable users to access multiple levels of detail in various regions simultaneously within ideograms.
The superiority of a particular method depends on the specific task at hand. Our findings indicate that \cVREmbedded{} was the fastest in summarizing gene phenotypes. On the other hand, \cPC{} outperformed the others in identifying gene distribution and comparing gene orientation, but it was ranked as the least preferred option by participants when compared to other VR conditions.
Interestingly, VR-Embedded also exhibited faster performance compared to VR-Insets in identifying gene distribution and summarizing gene phenotypes. This suggests that there are some advantages to placing detailed information directly within the ideogram (embedded) rather than providing it at a distance (in insets) within the VR environment.
Overall, our results contribute to an empirical understanding of multi-focus interactions in both VR and desktop environments, specifically in the context of querying human genome information. 
These findings also shed light on novel multi-focus interactions.

\bibliographystyle{abbrv-doi}

\bibliography{template}
\end{document}